\PassOptionsToPackage{round,sort&compress}{natbib}
\PassOptionsToPackage{hidelinks}{hyperref}

\documentclass[pdflatex]{sn-jnl}

\usepackage{graphicx}
\usepackage{amsmath,amssymb,amsfonts}
\usepackage{bm}
\usepackage{xcolor}
\usepackage[normalem]{ulem}
\usepackage{footmisc}
\usepackage{lineno}
\usepackage{multirow}
\usepackage{mathrsfs}
\usepackage[title]{appendix}
\usepackage{textcomp}
\usepackage{manyfoot}
\usepackage{booktabs}
\usepackage{algorithm}
\usepackage{algorithmicx}
\usepackage{algpseudocode}
\usepackage{listings}
\usepackage{siunitx}

\providecommand{\un}[1]{\si{#1}}        

\begin{document}
\title{Remanent crustal strain on Mars in non-poikilitic olivine of NWA 7721}

\author*[1,2]{\fnm{Yaozhu} \sur{Li}}
\author[3]{\fnm{Szilvia} \sur{Kalácska}}
\author[2]{\fnm{Phil} \sur{McCausland}}
\author[2]{\fnm{Roberta L.} \sur{Flemming}}
\author[4]{\fnm{Callum} \sur{Hetherington}}
\author[4]{\fnm{Bo} \sur{Zhao}}
\author[1]{\fnm{Can} \sur{Yildirim}}
\author[1]{\fnm{Carsten} \sur{Detlefs}}

\affil*[1]{\orgname{European Synchrotron Radiation Facility}, \orgaddress{\street{71 Av. des Martyrs}, \city{Grenoble}, \postcode{38000}, \country{France}}}

\affil[2]{\orgdiv{Department of Earth Sciences}, \orgname{Western University}, \orgaddress{\street{1151 Richmond Street}, \city{London}, \state{Ontario}, \postcode{N6A 5B7}, \country{Canada}}}

\affil[3]{\orgdiv{Mines Saint-Etienne}, \orgname{Univ Lyon, CNRS, UMR 5307 LGF, Centre SMS}, \orgaddress{\street{158 cours Fauriel}, \postcode{42023}, \city{Saint-Étienne}, \country{France}}}

\affil[4]{\orgdiv{Department of Geosciences}, \orgname{Texas Tech University}, \orgaddress{\street{2500 Broadway W}, \city{Lubbock}, \postcode{74909}, \state{Texas}, \country{United States}}}

\abstract{
We present a multiscale microstructural analysis of olivine from the non-poikilitic lithology of the poikilitic shergottite NWA 7721, using dark-field X-ray microscopy (DFXM), electron backscatter diffraction (EBSD) and context in situ 2D X-ray diffraction (µXRD). A single olivine crystal contains two distinct subgrain populations. Type 1 subgrains are fine (1--5\un{\mu m}), randomly oriented, and nearly strain-free, whereas Type 2 subgrains are coarse ($>$30~$\mu$m), aligned, and strongly strained. Layered DFXM data reveal slip-band features in Type 2 that are absent in Type 1. We interpret Type 1 as products of shock-induced recrystallization, whereas Type~2 preserve remnants of a highly deformed parent grain. This bimodal microstructure, not observed in other Martian meteorites including the paired NWA 1950 and ALH A77005, points to a heterogeneous response to impact, influenced by pre-existing strain of the olivine grain. We propose that NWA 7721 olivine had already experienced substantial crustal or magmatic stress before impact. The subsequent shock wave imposed a rapid load--release cycle that mobilized dislocations and produced low-angle boundaries in Type 2, while driving recrystallization of Type 1. Grain-growth constraints limit the post-shock heating duration to $\approx 2.3\un{s}$, consistent with rapid quenching. These results provide the first evidence that non-poikilitic olivine in NWA~7721 preserves dynamic crustal deformation on Mars in the late Amazonian time.}

\maketitle

\section{Introduction}
Martian shergottites are the most abundant martian meteorites and record young Martian magmatic activity and impact histories \cite{nyquist1998shergottite,borg2002constraints,mcsween1998martian, combs2019petrology}. Poikilitic and gabbroic shergottites are emplaced deep in the Martian crust, separating them from basaltic and olivine-phyric shergottites that represent a hypabyssal to extrusive origin \cite{combs2019petrology,mcsween1998martian, gillet2005petrology}. Poikilitic shergottites ($\approx20 \%$ of Martian meteorites) are petrographically unique because of their bimodal texture, composed of poikilitic lithology with pyroxene oikocrysts enclosing olivine grains, and non-poikilitic lithology exhibiting olivine and pyroxene cumulates. A two-stage polybaric origin is proposed: chamber crystallization of poikilitic domain followed by degassing-assisted ascent forming non-poikilitic olivine \cite{howarth2014two, combs2019petrology}. Olivine in the non-poikilitic domain thus may have been exposed to the evolving magmatic stress fields before subsequent impact loading, offering a rare insight into how Martian mantle-derived magmas accommodated deformation during cooling and how subsequent shock events modified these early strain signatures. 

Olivine develops distinct microstructures under varying deformation regimes: it develops lattice-preferred orientations (LPO) and slip systems via dislocation glide \cite{Mainprice2005, Raterron2007, Ohuchi2011} at low strain rates ($\approx 10^{-15}$/s), whereas at hyper-velocity impacts with extreme strain rates (up to $10^4$/s), olivine produces disordered crystallographic plasticity such as shock mosaicism \cite{fritz_revising_2017, vinet2011crystal,stoffler2018shock,li2020best, flemming2007micro, li2025petrology}. However decoding the entangled non-shock and shock textures remains challenging for reconstructing the mechanical and thermal evolution in meteoritic materials. 

In this study, we combine multi-scale analysis of in-situ Two-dimensional micro X-ray diffraction ($\mu$XRD)(2D µXRD), electron backscatter diffraction (EBSD) and synchrotron-based dark-field X-ray microscopy (DFXM) to investigate non-poikilitic olivine microstructure in shergottite Northwest Africa (NWA) 7721. We identify a striking bimodal subgrain microstructure within single petrographically contiguous olivine crystals. 

Fine-grained (3–-5\un{\mu m}), randomly oriented, nearly strain-free subgrains (Type 1) coexist with coarse ($>30\un{\mu m}$), highly strained subgrains (Type 2) that host continuous low-angle grain boundaries (LAGBs) forming slip-band-like features across multiple $z$-layers. The observed microstructure cannot result from a shock event alone. Instead, it records the overprint of shock-induced dynamic recrystallization onto pre-existing, dislocation-rich domains. DFXM analysis of orogenic olivine confirms that the patterned LAGBs in Type 2 reflect inherited slip patterns from crustal strain. We interpret Type 1 as products of shock-assisted recrystallization and Type 2 as relicts of shock and pre-shock tectonic loading, originated from the degassing event of the arising magma. Their coexistence within one crystal hints a unique thermomechanical pathway: a directional shock propagated along pre-existing stress fields. The internal back-stress field facilitated the rapid dislocation motion forming Type 2 subgrains while localized heating further enabled Type 1 nucleation. The observation of the pre-shock dislocation in Type 2 relicts also manifests a strong regional tectonic activity at where the shergottite magma was emplaced, illustrating an active Mars at least around 600\un{Ma} ago \cite{nyquist1998shergottite, cohen2023synchronising, herd2024source}.

Our results provide the microstructural evidence for the polybaric theory for the formation of the non-poikilitic lithology in the poikilitic shergottite. We bridge 2D crystallographic microstructure with 3D internal structure, enabling the distinction between shock-induced and inherited features. Our work expands the application of DFXM to geological materials, highlighting the value of a multi-scale microstructural approach in resolving complex deformation histories in planetary materials.

\section{Results}
\subsection{Bimodal internal morphology of non-poikilitic olivine}
\begin{figure}
    \centering
    \includegraphics[width=1\linewidth]{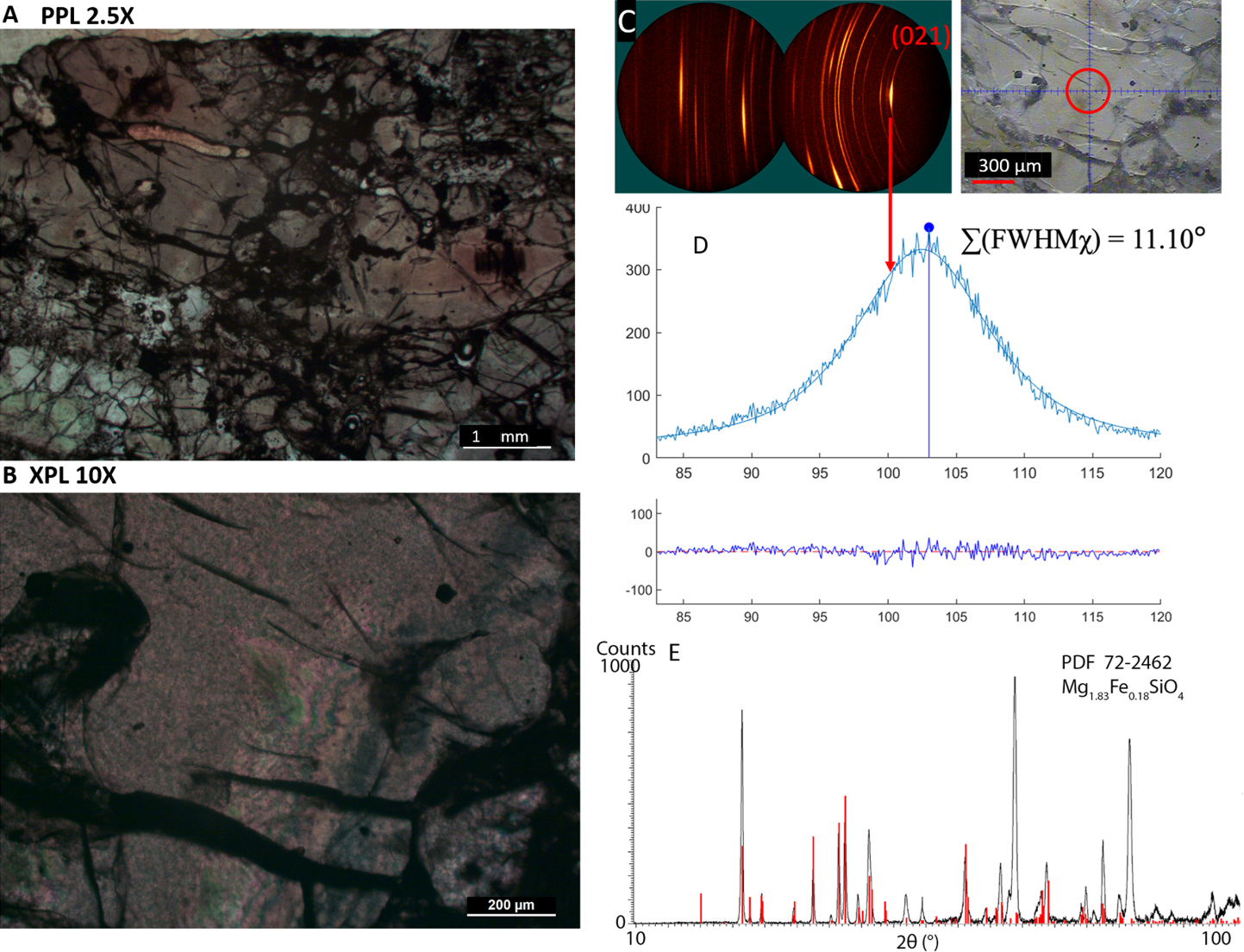}
    \caption{Macroscopic textures of NWA 7721. \textbf{A}: Context petrographic image PPL for investigated non-poikilitic olivine. \textbf{B}: 10X image XPL for investigated non-poikilitic olivine showing fringing  \textbf{C}: 2D XRD pattern showing strain-related diffraction streaks and polycrystalline rings along the Debye rings, with targeted in situ X-ray 300 µm diameter beam spot shown in context image of grain at right. \textbf{D}: $\sum(\mathrm{FWHM}_\chi)$ for the integrated peak at lattice plane (021). \textbf{E}: 1D XRD pattern along $2\theta$, matching with forsteritic olivine using Powder Diffraction File (PDF) from the International Center for Diffraction Data.}
    \label{figure1}
\end{figure}

The investigated non-poikilitic olivine has a homogeneous composition ($Fo_{69.62} \pm{1.26}$, N=8) surrounded by black mesostasis showing shock stain and mosaicity (Fig.~\ref{figure1}A and Fig.~\ref{figure1}B, SI).2D $\mu$XRD shows combined streaky diffraction and a faint powder-ring patterns along the Debye ring (Fig.~\ref{figure1}C). The quantitative strain-related mosaicity (SRM) using best-fit-for-complex-peaks (BFCP) analysis \cite{li_quantitative_2021, li2020best, flemming2007micro, vinet2011crystal, Li2023} yields $\sum(\mathrm{FWHM}_\chi)$ of $11.10^ \circ$ (Fig.~\ref{figure1}D). The integrated 1D XRD pattern of $2\theta$ fits well with a single phase olivine (Fig.~\ref{figure1}E) only, consistent with the coexistence of two XRD pattern textures within the single olivine grain. 

\begin{figure}
    \centering
    \includegraphics[width=1\linewidth]{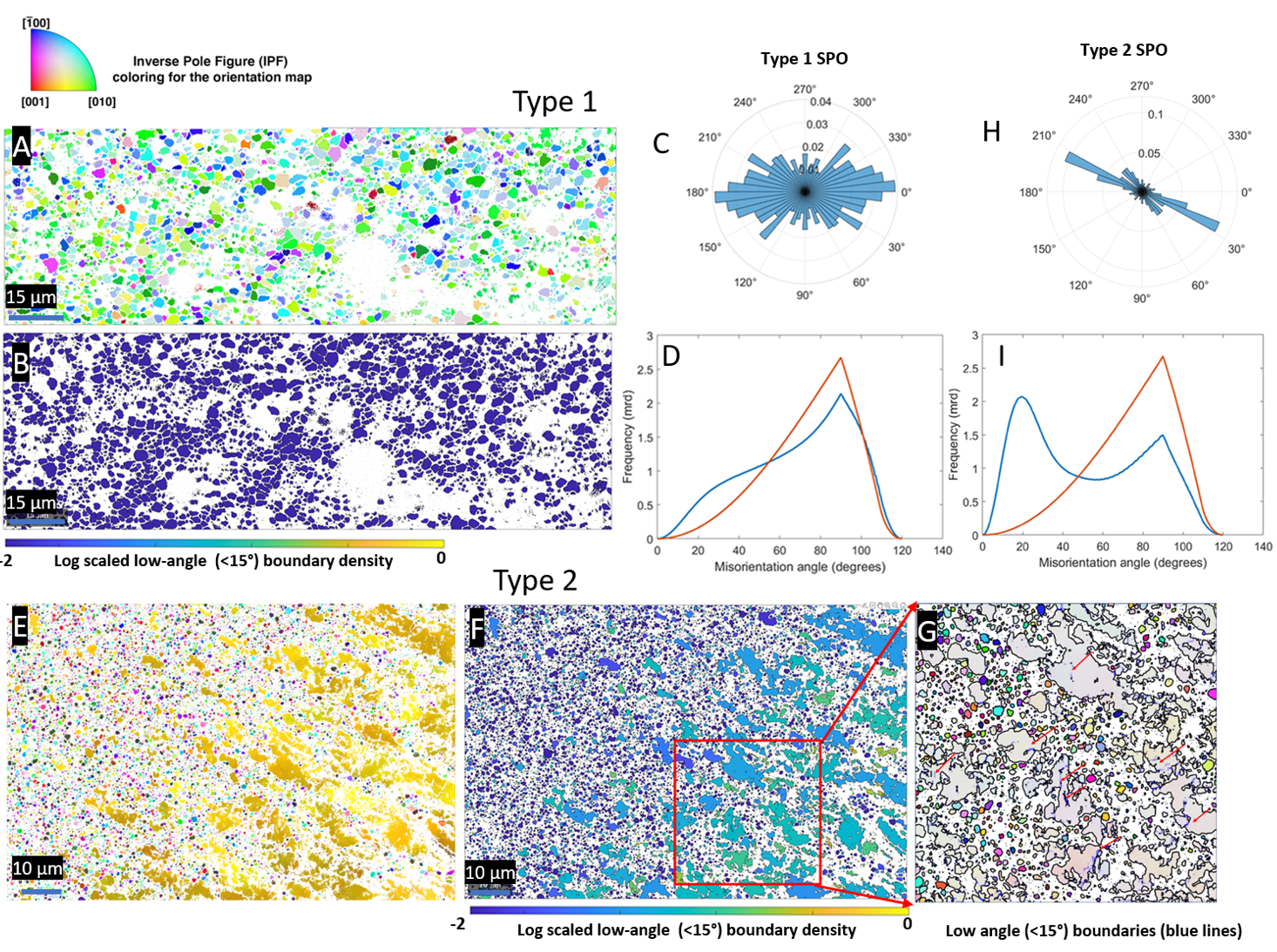}
    \caption{EBSD of Type 1 and Type 2 olivine subgrains. \textbf{A}: IPF-colored orientation plot of Type 1 subgrains. \textbf{B}: log scaled low-angle boundary (LAGB) density within each subgrain. \textbf{C}: shape-preferred orientation of Type 1 subgrains. \textbf{D}: misorientation angle distribution plot for Type 1 subgrains compared with a theoretical random distribution (red line). \textbf{E}: IPF-colored orientation plot showing both Type 1 subgrains and Type 2 subgrains. \textbf{F}: log scaled LAGB density for Type 2 subgrains. \textbf{G}: LAGBs (blue lines) in Type 2 subgrains. \textbf{H}: Shape preferred orientation for Type 2 subgrains showing strong alignment. \textbf{I}: misorientation angle distribution plot compared with a theoretical random distribution (red line)}
    \label{figure2}
\end{figure}

EBSD mapping reveals that the macroscopically continuous olivine single grain is internally partially recrystallized. Fine-grained aggregates (3--5\un{\mu m}) show nearly random orientation(Fig.~\ref{figure2}A) with minimum low-angle boundaries (LABs, misorientation $<15^\circ$, Fig.~\ref{figure2}B), no shape-preferred orientation (SPO, Fig.~\ref{figure2}C), and nearly random misorientation angle distribution (Fig.~\ref{figure2}D) \cite{Skemer2005}. These subgrains are classified as Type 1. In contrast, we found regions within the same grain containing bimodal morphologies of Type 1 subgrains coexisting with large irregular subgrains (15\un{\mu m}), showing a strong preferred crystallographic orientation (Fig.~\ref{figure2}E), the development of LAGBs (Fig.~\ref{figure2}F and \ref{figure2}G), and strong SPO (Fig.~\ref{figure2}H). These large irregular subgrains are classified as Type 2. The misorientation angle profile in this region shows a random distribution for Type 1 subgrains and a strong preferred distribution around $20^\circ$ (Fig.~\ref{figure2}I) for Type 2 subgrains.

\subsection{Dark-Field X-ray Microscopy}
\begin{figure}
    \centering
    \includegraphics[width=0.95\linewidth]{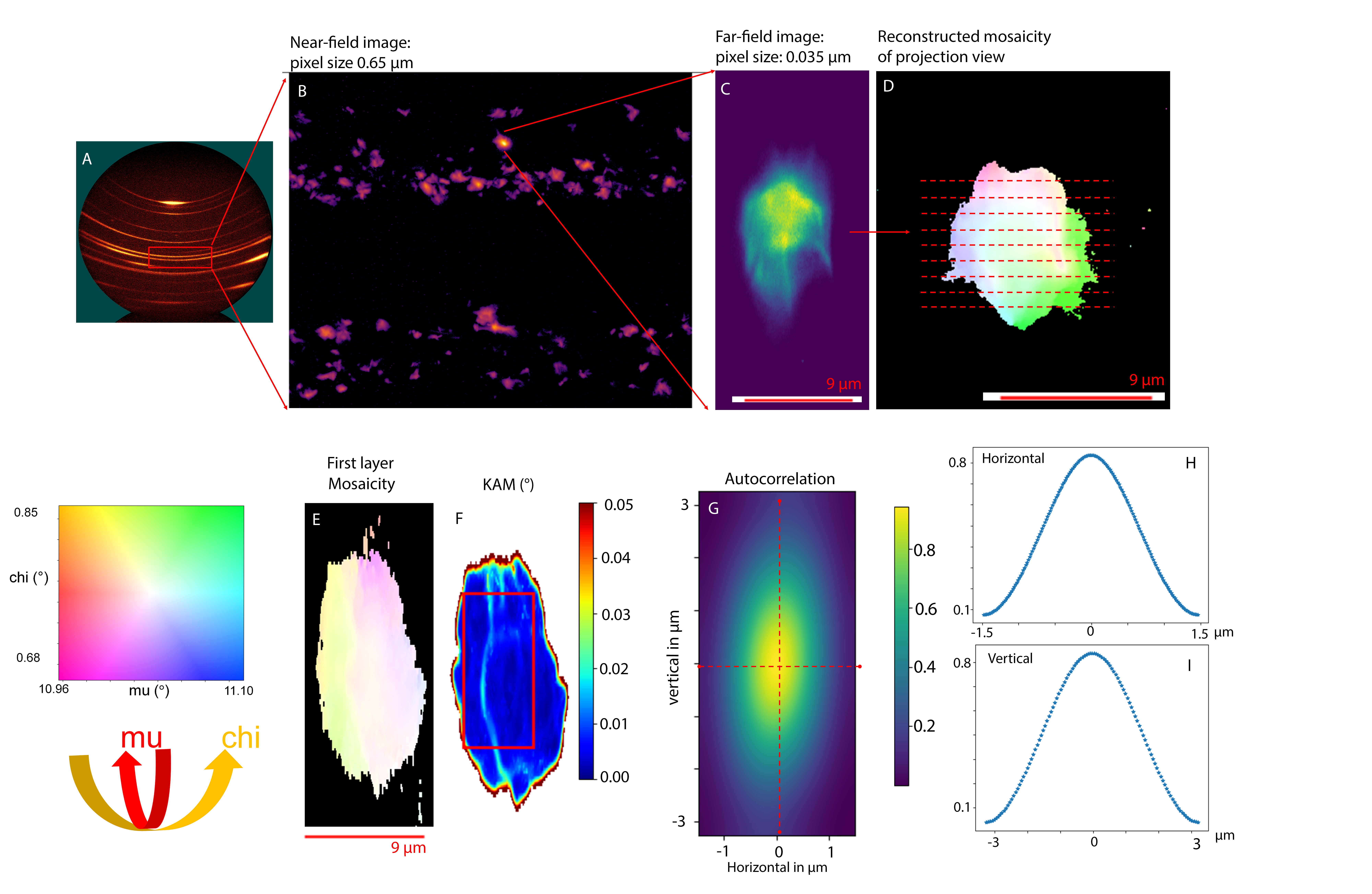}
    \caption{Dark-field X-ray microscopy for Type 1 subgrains. \textbf{A}: Example of 2D XRD image of the investigated grain. \textbf{B}: Zoom-in view of lab 2D XRD showing the powder ring pattern using near-field camera with 2500 pixels horizontal and 2000 pixels vertical. \textbf{C}: Zoom-in view of selected subgrain imaged by the far-field detector with an objective lens. \textbf{D}: Reconstruction of the projection of the selected subgrain using $100 \times 100 \un{\mu m}$ beam size. Dash lines are schematic demonstration of using a line-focus beam (with $500\un{nm}$ beam height of $1\un{\mu m}$ spacing) to image different layers in the grain. \textbf{E}: Reconstruction of the first layer mosaicity showing small mosaic spread \textbf{F}: First layer Kernal Average Misorientation map highlighting the small misorientation boundaries \textbf{G}: Autocorrelation analysis of the selected area (red box in F). \textbf{H}: Horizontal 1D projection from G. \textbf{I}: Vertical 1D projection from G.}
    \label{figure3}
\end{figure}

\begin{figure}
    \centering
    \includegraphics[width=1\linewidth]{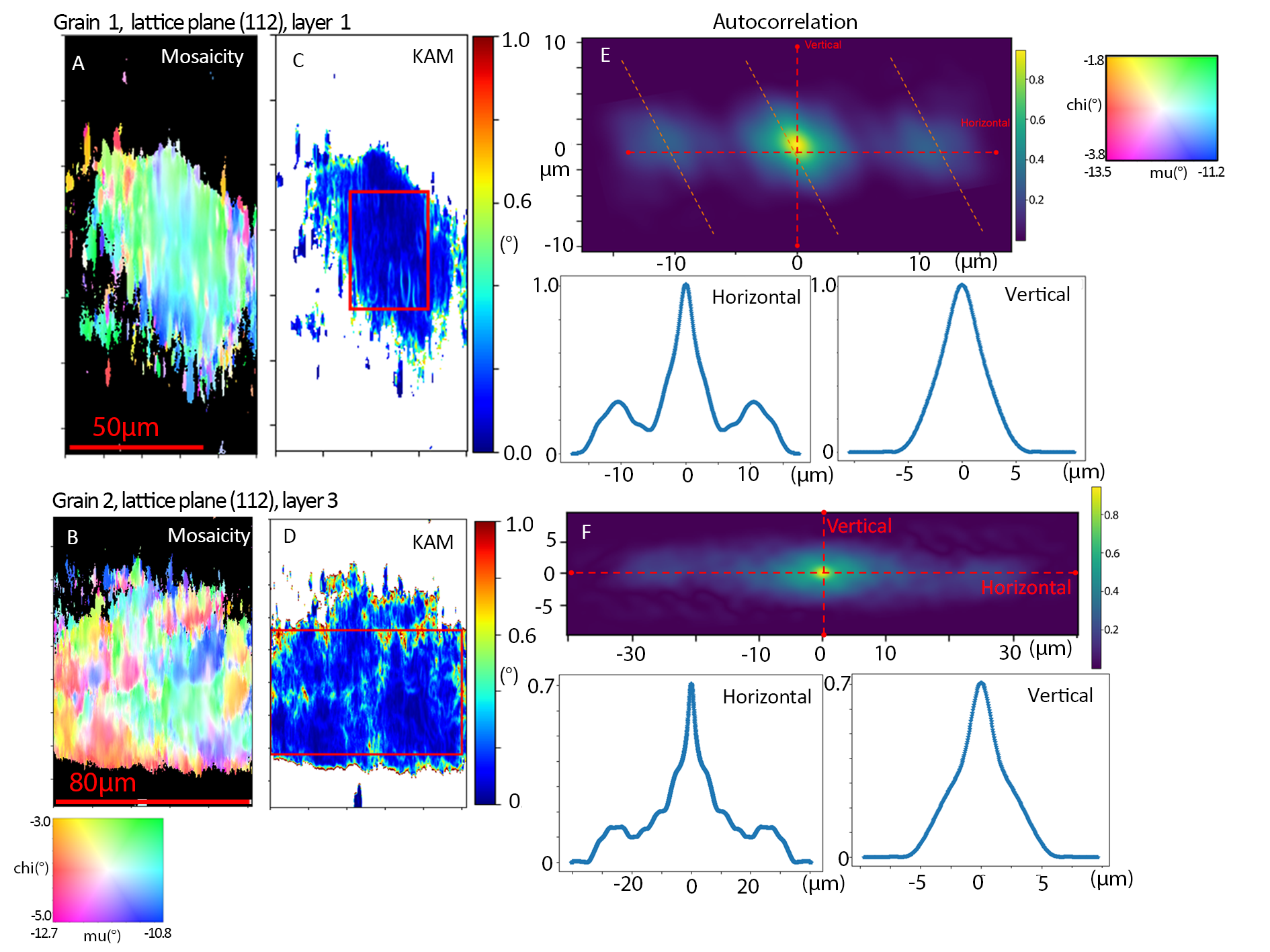}
    \caption{Dark-Field X-ray Microscopy for Type 2 subgrains. \textbf{A} and \textbf{B}: chi-mu mosaicity plots. \textbf{C} and \textbf{D}: Kernel average misorientation plots calculated from center of mass from mu motor. \textbf{E} and \textbf{F}: autocorrelation orientational analysis}
    \label{figure4}
\end{figure}

Similar ring patterns for Type 1 at (112) and (130) planes appeared in the near-field camera (Fig.~\ref{figure3}A). A high-resolution real-space image was obtain for one of subgrain reflections in the (112) ring (Fig.~\ref{figure3}B). Illumination with a $100 \times 100\un{\mu m^2}$ box beam shows the full projection of a 6\un{\mu m} subgrain (Fig.~\ref{figure3}D) and a discrete mosaicity spread (Fig.~\ref{figure3}E). Layering image (10 slices with 1\un{\mu m} spacing) showed a consistent small mosaicity of maximum chi ($\chi$) $0.17^\circ$ and mu ($\mu$)  $0.14^\circ$ spread (Fig.~\ref{figure3}E and F). A very low misorientation angle ($<0.02^\circ$)near subgrain edge was observed on the kernel average misorientation (KAM) map (Fig.~\ref{figure3}G), consistent in each slices. In the investigation area (red box), autocorrelation analysis shows a homogeneous orientation spread with a smooth Gaussian 1D projection (Fig.~\ref{figure3}H), confirming its strain-free nature. 

Type 2 subgrains were imaged with a coarser spacing (5\un{\mu m}) for 10 layers due to their large size and prolonged collection time. They exhibit a large mosaicity with $\chi$ and $\mu$ around $2 ^\circ$ (12 times larger than Type 1) (Fig.~\ref{figure4}A and \ref{figure4}B). The imaged subgrain boundary misorientation is up to $1 ^\circ$ (20 times larger than Type 1). Within each layer, we observed the aligned dislocations forming subdomains with the small misorientation angle around $0.5^\circ$ (Fig.~\ref{figure4}C and \ref{figure4}D), persisting through the grain, including the occasional ``triple-junction'' feature formed at subdomain intersections (Fig.~\ref{figure4}D). We also observed less aligned boundaries among subdomains formed by very-low-angle misorientation boundaries (< $0.1^\circ$ to $0.2^\circ$), connecting them into a dislocation network (Fig.~\ref{figure4}C and D). Autocorrelation analysis reveals a repeating orientational pattern with a spacing of 10 to 25\un{\mu m} superimposed on the broadened Gaussian curves (Fig.~\ref{figure4}E and \ref{figure4}F), throughout all investigated layers.

\begin{figure}
    \centering
    \includegraphics[width=1\linewidth]{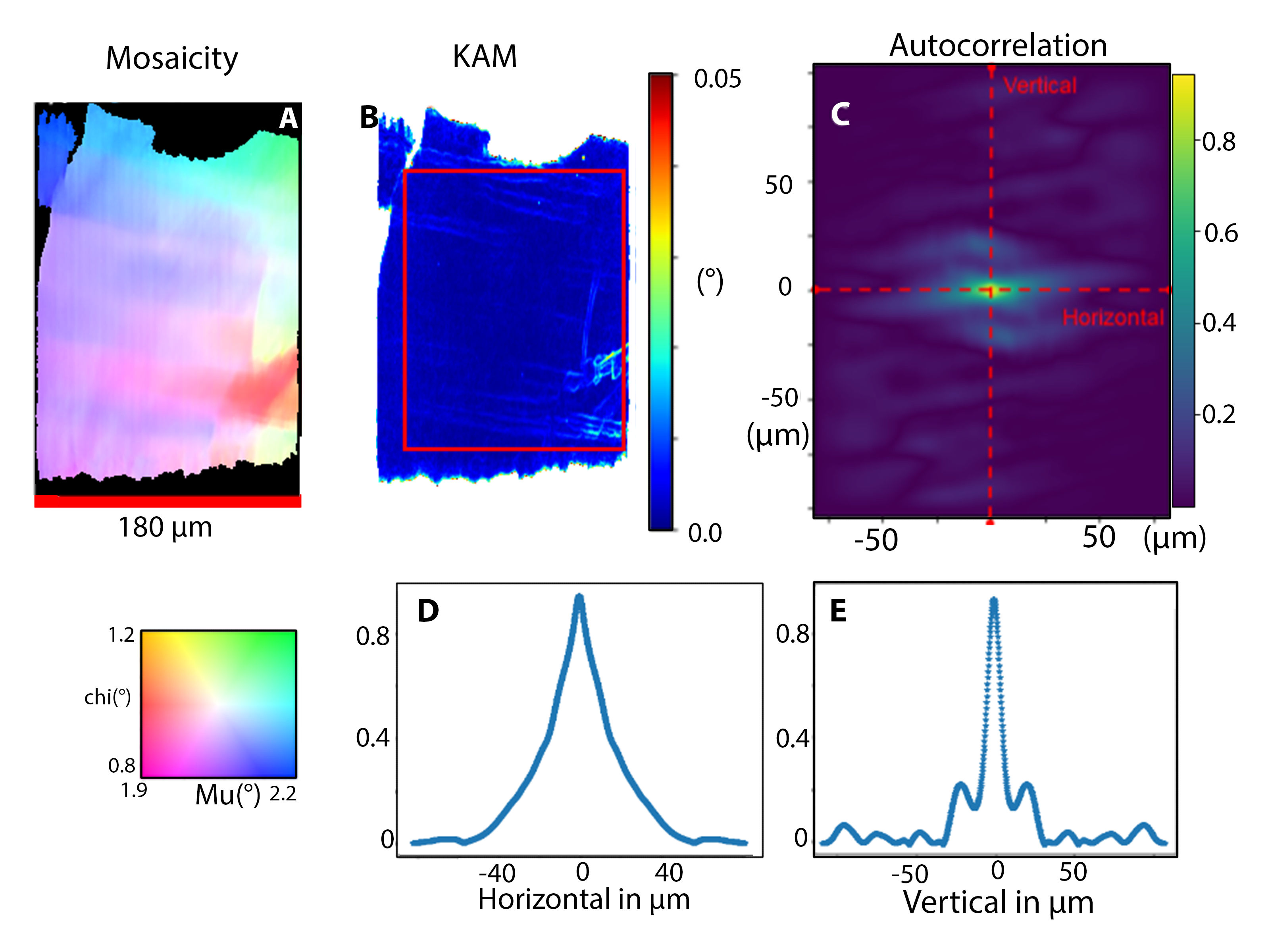}
    \caption{Dark-field X-ray microscopy for {\AA}heim olivine. \textbf{A}: Mosaicity plot of $\chi-\mu$. \textbf{B}: KAM plot. \textbf{C}: autocorrelation orientational analysis. \textbf{D}: 1D horizontal projection of autocorrelation analysis. \textbf{E}:1D vertical projection of autocorrelation analysis.}
    \label{figure5}
\end{figure}

To verify the DFXM observations of the slip-band feature in Type 2 subgrains, a similar analysis was performed on an {\AA}heim olivine peridotite from Norway. The {\AA}heim peridotite is found to have Type B ([001](010)) and A ([100](010)) LPO with Type E ([100](001)) slip \cite{jung2021dislocation} for subgrain boundaries. We observed the mild mosaicity (Fig.~\ref{figure5}A)(max spread $\chi$~$0.4^\circ$ and $\mu$ ~ $0.3^\circ$). The KAM map revealed the alignment of the incipient subdomain boundaries with a very low misorientation angle of $0.02^\circ$. The autocorrelation analysis further revealed a slip-band feature evidenced by the narrow discrete repeating curves with a spacing of 20 to 25\un{\mu m} (Fig.~\ref{figure5}C, Fig.~\ref{figure5}D, and Fig.~\ref{figure5}E) originated by tectonic shear. 

\section{Discussion}
\subsection{Impaired deformation microstructure in olivine single crystal}
\begin{figure}
    \centering
    \includegraphics[width=1\linewidth]{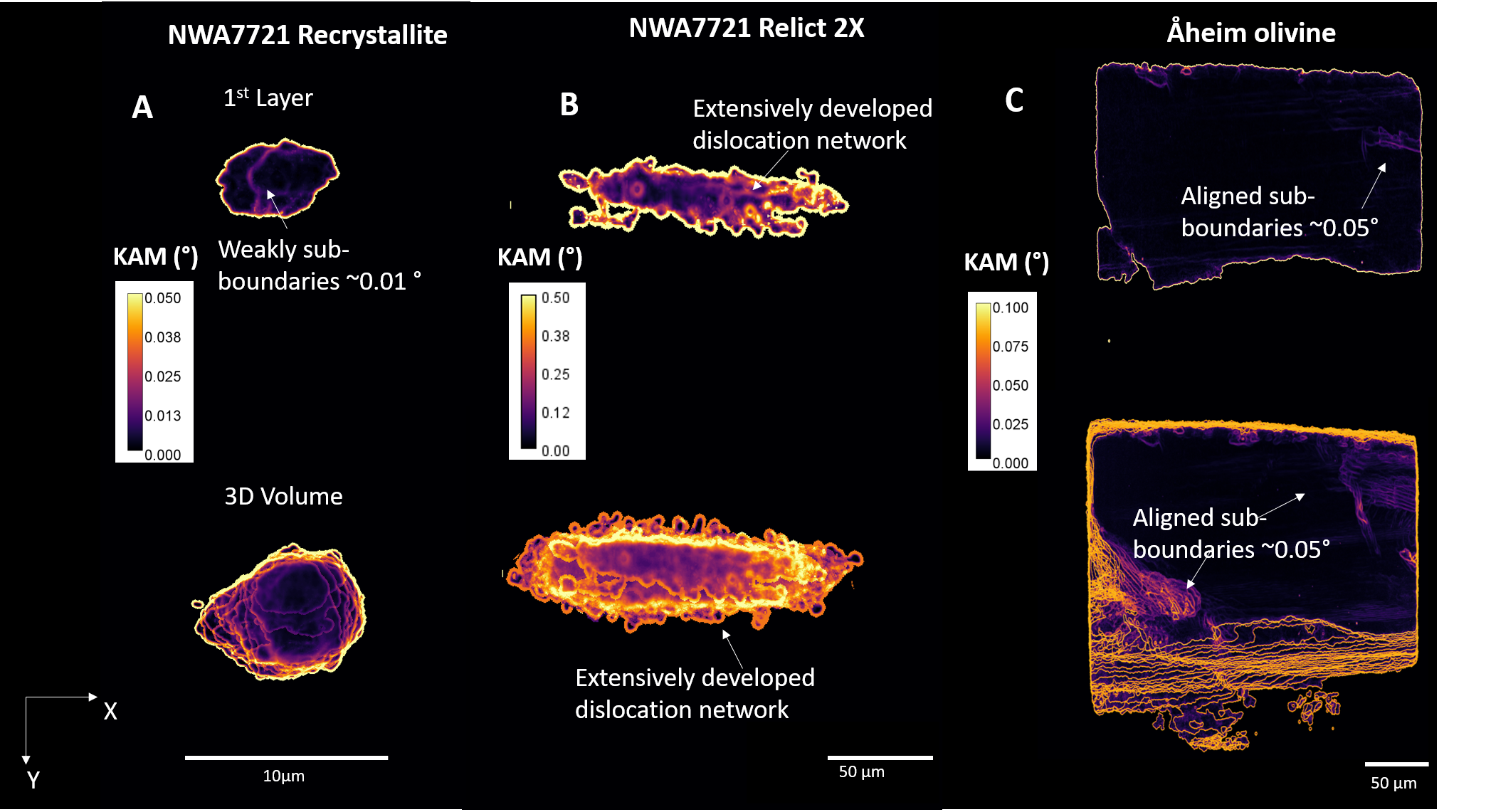}
    \caption{DFXM 3D reconstruction highlighting the developed sub-boundaries. \textbf{A}: NWA 7721 recrystallites (Type 1 subgrain). \textbf{B}: NWA 7721 relict subgrain (Type 2 subgrain). Top row is the view from Z-axis showing one layer, and bottom row is view rotation along Y-axis. \textbf{C}: {\AA}heim olivine. The full video is in the supplementary material.}
    \label{figure7}
\end{figure}

To the best of our knowledge, this is the first report of a bimodal microstructure with full 3D reconstruction for Martian poikilitic shergottite (Fig.~\ref{figure7}A and Fig.~\ref{figure7}B). 

The 2D-XRD patterns shows heterogeneous mosaic block sizes and orientation in $300 \times 300\un{\mu m}$ area. Long streaks indicate misoriented mosaic blocks ($< 15 \un{\mu m}$) caused by non-uniform strain whereas the powder-ring pattern along the Debye rings reflect even smaller mosaic blocks ($< 5 \un{\mu m}$) with nearly random orientation \cite{li2020best, li_quantitative_2021, Li2023, vinet2011crystal, flemming2007micro, horz1973debye, li2023witness, li2025petrology}. EBSD further confirmed that Type 1 subgrains have uniform small size, randomized orientation, and minimum accumulation of plastic strain, implying recrystallization rather than annealing; while Type 2 subgrains show the preferred orientation with the extensive development of LABs. The coexistence of both textures within a single grains is inconsistent with the microstructure formed solely by a transient shock event, suggesting the possible contribution from both magmatic strain and impact metamorphism. 

Previous studies suggest poikilitic shergottite was deformed at a high temperature of $1750$ -- $1870\un{K}$ with pressure of 55\un{GPa}, inferred from the shock melt and brown olivines \cite{takenouchi_mineralogical_2017}. Theses conditions facilitate olivine nucleation and growth by enhancing the dislocation mobility within the strained crystal lattice. Applying grain growth law (see SI) with established experimental studies\cite{nichols1991grain,speciale2020rates,karato1989grains, faul2006grain}, recrystallizing Type 1 subgrain requires a very brief post-shock heating (2.3s) within an $1 \un{mm}$ domain at $1800\un{K}$, consisting with the brief shock pulse duration and the fast but limited growth of Type 1 subgrain. 

However, the same condition cannot form the larger strained Type 2 subgrains. In olivine, the development SPO accompanied by LPO generally reflects the long-lived deformation process of tectonic or magmatic strains \cite{zaffarana2014microstructures,jung2021dislocation,karato_karato_2003,hansen2014protracted,hansen_low-temperature_2019}. Without sufficient post-shock recovery, a single shock pulse with seconds timescale alone is unlikely to form aligned fabrics \cite{fritz_revising_2017, stoffler2018shock}. Moreover, no systematic SPO or LPO have been reported in the discovered Martian meteorites \cite {Treiman2007, tkalcec2019absence, combs2019petrology, fritz_revising_2017}. Therefore, we interpreted Type 2 subgrains are relicts of pre-shock olivine, and the observed SPO as an inherited orientation from its deformed parent grain, while the Type 1 subgrains represent shock-assisted recrystallization.    

\subsection{Convoluted microstructures in the relict subgrains}
Besides SPO, the Type 2 subgrain further shows the dense dislocation network composed by very-low-angle boundaries, and we propose these as the hybrid microstructures of pre-shock and shock plastic deformation. Experiment work on olivine shows that load-unload deformation cycle can generate strain hardening, internal back-stress, and a Bauschinger effect that facilitate yielding on the reverse loading stage\cite{hansen_low-temperature_2019}. 

The dense dislocation network is only observed in Type 2 subgrains, concluding these very-low-angle boundaries was not generated in the post-shock stage. As the preserved fragments of the parent crystal, we assumed that the Type 2 subgrains was deformed in a simple-shear flow-like deformation at the late-stage magma ascending when forming non-poikilitic olivine. With the SPO of misorientation $20 ^\circ$, we estimated a finite non-coaxial bulk strain of $\gamma \sim 0.36$ \cite{ramsay_techniques_1984}. This corresponds well with the incremental shear strain accommodated by dislocation glide forming the aligned very-low-angle boundaries with 10 to 25 \un{\mu m} spacing within a $\Delta\gamma \sim 0.036 - 0.0144$ radians, corresponding to $2.1^\circ - 0.82 ^\circ$ \cite{ashby1970deformation}. This observation consistent with the experimental progressive subgrain rotation and fabric development in olivine under high-temperature shear \cite{hansen2014protracted}.

This estimated magnitude of intracrystalline strain is consistent with KAM from DFXM layer reconstruction (Fig.~\ref{figure7}), where the aligned subboundaries with larger misorientation angle of $0.5^\circ - 1 ^\circ $ contributed to the repeating orientation patterns on the autocorrelation analysis (Fig.~\ref{figure7}B and Fig.~\ref{figure7}C). The shock-induced misorientation has angle of $ <0.5^\circ$ (Fig.~\ref{figure7}A and Fig.~\ref{figure7}B), randomly distributed between the subboundaries, contributed to the broadening on the autocorrelation analysis. 

The {\AA}heim olivine from tectonic shear shows the similar periodic banding but narrow autocorrelation peaks. The mosaic spread from $\chi-\mu$ in {\AA}heim olivine is twice that of Type 1 subgrains but almost four times smaller than Type 2 subgrain. Together these observations concludes Type 2 records a pre-existing microstructural fabric with shock-induced disorder while Type 1 is shock-assisted and nearly strain free recrystallites. 

\subsection{Preservation of entangled microstructure}
The polybaric crystallization for poikilitic shergottites hypothesize the non-poikilitic grains was formed during the ascent of magma towards the surface\cite{combs2019petrology, howarth2014two}. The the aligned LAGBs with repeating orientation pattern revealed in the Type 2 subgrains reflects a pre-shock fabric that was partitioned and partially overprinted during impact, rather than created newly.

We propose a three-stage process for the formation/preservation for these microstructures: (1) \textit{Pre-shock:} Late-stage ascent/degassing imposed localized simple-shear–like differential stresses, producing a strong internal orientation gradients and high geometrically necessary dislocations (GNDs) density in the parental crystal. The GNDs form the aligned LABs producing the junctions and walls established long-range internal stresses. (2) \textit{Shock loading and unloading:} Compression shock wave acted along this grains. Upon rapid unloading, internal back-stress facilitates the reverse yielding and the dislocation motion \cite{hansen_low-temperature_2019}), instantaneously partitioned the parent fabric into irregular subdomains bounded by LAGBs. (3) \textit{Post-shock retention:} The brief post-shock heating within seconds limited recovery and further boundary migration, preserving: (i) fine, nearly strain-free Type 1 recrystallites formed near melt, and (ii) Large Type 2 relic domains with \(\sim0.5^\circ\!-\!1^\circ\) LAGBs, \(\sim20\!-\!25~\mu\mathrm{m}\) banding, and incipient triple junctions that were partitioned from parent grain (Fig.~\ref{figure5}D). Thus the pre-existing  magma ascent-related fabric that was subsequently preserved and modified,but not erased, by shock.

\subsection{Coherent multiscale observation for shock-modified olivine}
Many shergottites likely share an ejection event at ca.3.0 Ma \cite{herd2024source}, and $^{40}$Ar/$^{39}$Ar ages of $161\pm9$ to $540\pm63$~Ma indicate sourcing from relatively young Martian volcanism \cite{cohen2023synchronising}. NWA 7721 is paired with NWA 1950 \cite{ruzicka2015meteoritical}, yet its olivine preserves a distinctive bimodal texture: shock-produced, nearly strain-free Type 1 recrystallites along with shock-modified Type 2 relic subgrains within the same crystal. Macroscopic staining, mosaicism and local melt along with SRM analysis place it near S5 shock stage\cite{li_quantitative_2021}. Grain-growth kinetics constrain post-shock heating to $\sim$2.3~s, too short for full recovery, allowing pre-existing dislocation networks to persist and to be partitioned rather than erased. The absence of pervasive high-pressure phases with only localized melt supports a moderate-intensity, short-duration shock applied to a pre-strained but relatively cold target, distinguishing NWA 7721 from cases such as ALH A77005\cite{walton2007localized}.

Our multiscale observations establish a coherent link from optical textures and 2D XRD to EBSD and 3D DFXM microstructures. EBSD shows the apparent “mosaicism” composed by fine, nearly strain-free Type 1 recrystallites and large Type 2 relic subgrains with strong SPO and low-angle boundary network, embedded within a macroscopic single crystal. DFXM de-convolutes these observations in volume: Type 1 subgrains are \(\sim 6\!-\!7~\mu\mathrm{m}\) and bounded by very-low-angle (\(<0.05^\circ\)) interfaces whereas Type 2 subgrains (\(>\!15\!-\!30~\mu\mathrm{m}\)) exhibit \(\sim 0.5^\circ\!-\!1^\circ\) LAGBs that define subdomains and \(\sim 20\!-\!25~\mu\mathrm{m}\) banding. The very-low-angle (\(<0.1^\circ\)) networks broaden autocorrelation peaks, distinguishing shock-related disorder from the periodic banding seen in the Åheim analogue.

This study highlights the value of multi-scale observations for decoding complex meteoritic deformation. The successful application of DFXM delivers non-destructive, volumetric reconstructions of microstructure, complementing the conventional optical and EBSD analyses. Collectively, the multi-scale observations in our case suggests that NWA 7721 retains a record of olivine deformation in shock and in the Martian crust as recently as 130–600\un{Ma} \cite{hartmann2001cratering,shih1982chronology,herd2024source}.

\section{Methods}
\subsection{Dark Field X-Ray Microscopy}
To further investigate the microstructures in these subgrains, we applied Dark-Field X-ray Microscopy (DFXM) at beamline ID03 of the European Synchrotron Radiation Facility (ESRF). We report the first application of this method to highly shocked geological materials, expanding on its conventional use in materials science investigation of microstructures in metals, alloys, and ceramics. The Dark-Field X-Ray Microscopy experiments were performed at the ESRF beamline ID03. Dark Field X-ray Microscopy (DFXM) combines X-ray (section) diffraction topography and microscopy \cite{Isern2024, Simons2015, Poulsen2017, Poulsen2021}. By placing an X-ray objective lens into the Bragg diffracted beam between the sample and a high-resolution imaging detector, it allows the reconstruction of the real-space image from the reciprocal pattern \cite{Simons2015}. Compared to classical near-field topography, this objective lens not only magnifies the image, but its numerical aperture ($\approx 0.5\un{mrad}$) also defines the angular resolution of the diffracted beam \cite{Poulsen2017}. Being a diffraction technique, DFXM is sensitive to variations of the crystal lattice \cite{Poulsen2021}. Our experiment was carried out with a synchrotron stored current of 198\un{mA}. X-rays with photon energy 17\un{keV} were selected by a Double Multilayer Monochromator (DMM) followed by a Channel-cut Crystal Monochromator (CCM, Si(111) ) with a bandwidth of $\Delta E/ E \approx 10^{-2}$ and $\Delta E/ E \approx 10^{-4}$, respectively. We selected olivine lattice plane (112) for this study. The monochromatic beam was focused in the vertical direction by a Compound Refractive Lens (CRL) condenser comprised of 58 bi-parabolic (1D focusing) Be lenslets with an R=100\un{\mu m} radius of curvature, yielding an effective focal length of 908\un{mm}. The beam profile on the sample was approximately $200\un{\mu m(h)} \times 0.6\un{\mu m(v)}$ (FWHM). This horizontal \textit{line beam} illuminated a single plane that sliced through the depth of the crystal, defining the microscope's \textit{gauge volume} or ``layer'', as shown in Fig.~\ref{figure3}. The sample could be moved through the line beam such that the layers could be stacked into a 3D dataset.
The entry plane of the objective CRL was positioned in the Bragg diffracted beam, 281\un{mm} downstream of the sample. 

The far-field imaging detector used an indirect X-ray detection scheme. This detector was comprised of a scintillator crystal to convert X-rays into visible light, a visible light microscope and sCMOS camera (PCO.edge 4.2bi with $2048 \times 2048$ pixels). It was positioned 5010\un{mm} downstream of the sample. The visible light optics inside the far-field detector could switch between $10\times$ and $2\times$ magnification to achieve an effective pixel size of 0.65\un{\mu m} or 3.25\un{\mu m} at the scintillator, and 36.3\un{nm} or 182\un{nm} at the sample position, respectively. 

Data were analyzed using \textit{darfix} \cite{Garriga2023}. Kernel average misorientation was calculated by using the center of mass of motor values for each pixel in the collected data. The autocorrelation was then performed by computing the 2D fast-Fourier transformation in each map through out the layers, similar to the method being applied to Al materials in the previous studies. \cite{zelenika20243d,zelenika2025observing}.  

\subsection{Electron Backscatter Diffraction and Energy Dispersive Spectroscopy}
NWA 7721 was polished with diamond paste and finished with colloidal silica on a vibropolisher for electron backscatter diffraction (EBSD) analysis. EBSD data for olivine was obtained at Texas Tech University, College of Arts Sciences Microscopy Laboratory. Data were collected using a Zeiss Crossbeam 540 FEG-SEM, with an Oxford Instruments silicon drift detector (SDD) Energy Dispersive Spectrometer (EDS) and EBSD camera complemented by AZTec software that integrates packages by HKL at Texas Tech University. The step size of 170\un{nm} was used to capture the deformed olivine. Additional EDS mapping and point analysis was done using a Zeiss Supra 55VP SEM equipped with an Oxford Instruments XMaxN 80 detector for detailed olivine and pyroxene composition determination. The composition is further calculated in Mg\#, which is defined as 100*Fo/(Fo+Fa) for olivine (Fo as forsterite and Fa as fayalite in mol), and 100*En/(En+Fs) for pyroxene (En for enstatite and Fs for ferrosillite in mol) . 

EBSD data for olivine was further plotted and analyzed in Matlab by the unit segment length (USL) method developed by \cite{li2023witness}. We modified the USL method, in detail, a three-step reconstruction approach was taken due to the original olivine crystals being broken down into smaller crystallite aggregates (2--3\un{\mu m}). We first computed the grain orientations to include large misorientation angles (up to $120^\circ$) to ensure that all crystallites in the mapping area were included in the analysis; then we reconstructed grains with boundary angle above $15^\circ$. Finally, we reconstructed subboundaries for misorientations below $15^\circ$. The codes were also designed to monitor changes in the boundary geometry using the angle of the misorientation axis and boundary trace. More details can be found in \cite{Li2023,li2023witness,Skemer2005}.

\subsection{2D X-ray diffraction}
In situ 2D X-ray diffraction data were collected with a Bruker D8 Discover µXRD in the Department of Earth Sciences at Western University (see Flemming \cite {flemming2007micro}  for more details). This work used theta-theta geometry and operated at 35\un{kV} and 45\un{mA} with a Co K$\alpha$ X-ray source (Co K$\alpha_1$: $\lambda$ = 1.78897\,\AA) and Göbel mirror parallel optics with pinhole collimator, producing a parallel beam with a nominal diameter of 300 µm. Diffracted X-rays were recorded on a Vantec-500 2D area detector using General Area Detector Diffraction System (GADDS) software obtaining 2D diffraction patterns (GADDS images) similar to Debye-Scherrer film. The 2D detected area is composed by one dimension of the radial 2 theta ($2 \theta$) whose angle increased from the beam and the other dimension of concentric chi angle ($\chi$) representing diffraction conditions met along the Debye ring for each $2 \theta$. Distance of the sample to detector is set as 12 cm, as the best compromise between angular resolution and detection area. X-ray diffraction data used in this work were collected by using Omega scan mode, where the 2-theta ($2 \theta$) angle remains constant while the X-ray source ($\theta_1$) and X-ray detector ($\theta_2$) are rotated simultaneously clockwise about an angle omega (where $\theta_1 + \theta_2 = 2 \theta$). To maximize signal-to-noise ratio for the collected 2D XRD pattern, we collected 1 hr for each frame for two frames for each target. 

Bruker AXS DIFFRAC plus Evaluation® (EVA) software was then used with reference to International Center for Diffraction Data (ICDD) database for mineral phase identification. The software, EVA, first integrated 2D XRD images into a 1D plot of intensity versus $2 \theta$ angle. The best mineral phase match in the ICDD datable was then used to identify the prospective mineral in the targeting area. For the quantitative strain-related-mosaicity (SRM) from 2D XRD images, we integrated the 2D patterns along $\chi$ direction, and we further measured the full-width half maximum (FWHM) of the integrated peak as $\sum(\mathrm{FWHM}_\chi)$ by the best-fit-for-complex-peaks (BFCP) developed by \cite{li2020best}. The quantitative SRM analysis has shown a close relationship of magnitude of  $\sum(\mathrm{FWHM}_\chi)$ with increasing shock stages \cite{li_quantitative_2021, li2020best, flemming2007micro, vinet2011crystal}. 

\section*{Acknowledgements}
We acknowledge the European Synchrotron Radiation Facility (ESRF) for the in-house beamtime for this work. We would like to thank H.~Isern, E.~Papillon and Th.~Dufrane for assistance in using beamline ID03. PJAM and RLF acknowledge ongoing research support from NSERC Discovery Grants. SK was funded by the French National Research Agency (ANR) under the project No. ANR-22-CE08-0012-01 (INSTINCT).

\clearpage
\appendix
\section*{Supplementary Information}
\renewcommand{\thefigure}{S\arabic{figure}}
\renewcommand{\thetable}{S\arabic{table}}
\renewcommand{\theequation}{S\arabic{equation}}
\setcounter{figure}{0}
\setcounter{table}{0}
\setcounter{equation}{0}
\section{Sample description}
\begin{figure}
    \centering
    \includegraphics[width=1\linewidth]{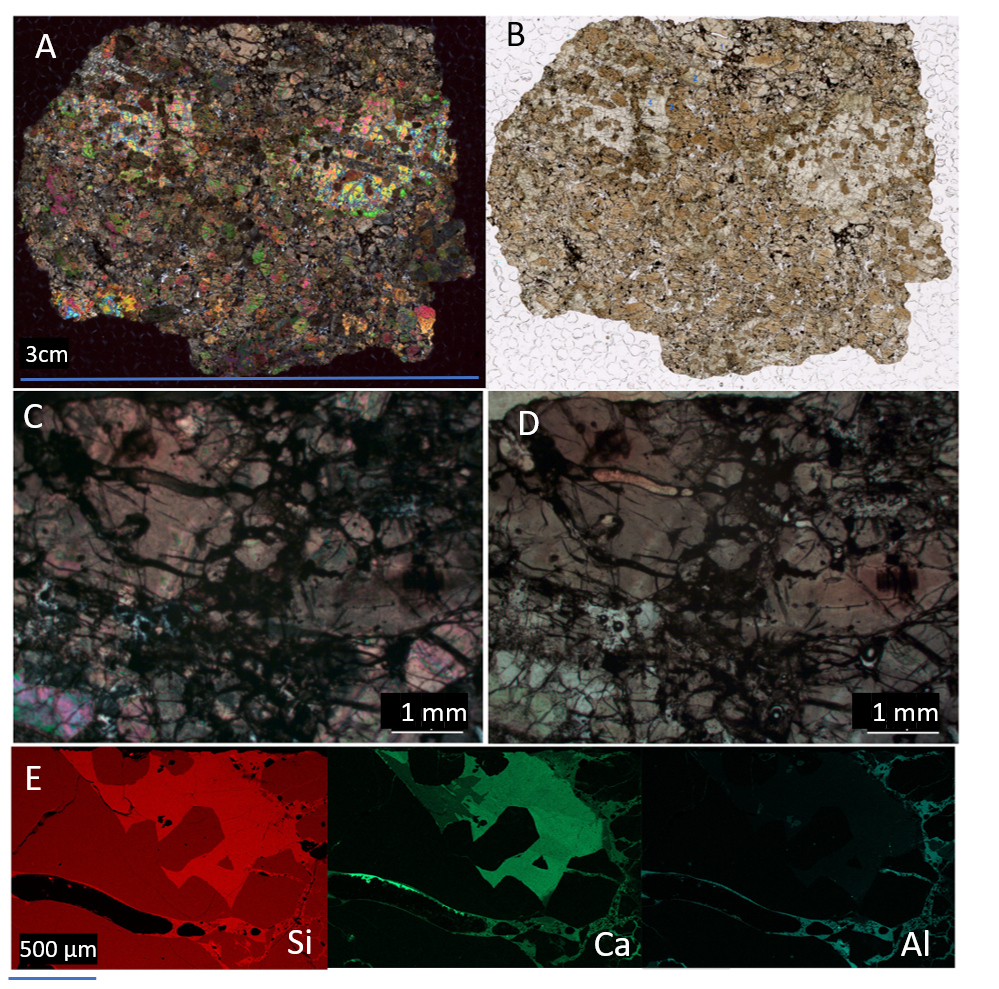}
    \caption{Petrographic images of NWA 7721\textbf{A}: Context petrographic image XPL for NWA 7721. \textbf{B}:  Context petrographic image PPL for NWA 7721 \textbf{C}: Petrographic image XPL in 2.5X for examined non-poikilitic olivine. \textbf{D}: Petrographic image PPL in 2.5X for examined non-poikilitic olivine. \textbf{E}: Energy dispersive scanning image of Si-Ca-Al highlighting metasomasis among olivine and clinopyroxene.}
    \label{figureS1}
\end{figure}

The investigated sample in our study is a standard thin section with 30 \si {\mu}m thickness (Fig.~\ref{figureS1}A and Fig.~\ref{figureS1}B). Olivine in non-poikilitic region shows cumulative textures, yielding a homogeneous composition ($\text{Mg}\#=69.62\pm{1.26},  N=8$), similar to non-poikilitic olivine observed in NWA 1950. Cracks, mosaicism, and shock stains are shown in these olivines (Fig.~\ref{figureS1}C and Fig.~\ref{figureS1}D). Augite (\textit{Wo}33.15) - pigeonite (\textit{Wo}11.35) are found adjacent to the olivine grains, and these clinopyroxenes yields a uniform Mg\# of  $73.34\pm{1.50},  N=5$. Black intracrystalline mesostasis is seen around olivine and clinopyroxene grains (Fig.~\ref{figureS1}B and \ref{figureS1}C). ESD mapping shows that these black metasomasis materials have similar compositional signature to their adjacent silicates of olivine but with higher in Si, Ca, and Al(Fig.~\ref{figureS1}E). 

\section{Grain growth}
A general grain growth law can be expressed as following: $[d^n -d_0^n = K_0\exp(-Q/RT)t]$, where $d^n$ and $d_0^n$ are the final and initial size of the crystal, respectively, $K_0$ is a scaling factor, $R$ is the gas constant, $T$ is the temperature, $n$ is the growth components of 2 to 4, and $Q$ is the activation energy from 200 to 600\si{kJ/mol} depending the presence of porosity and second phases. The microstructure of the Type 1 subgrains suggest a dynamic recrystallization instead of a static one, given its small crystal size, subhedral crystal shape, and random orientation.
The grain growth of the dynamic recrystallization, including both tectonism or hypervelocity impacts, is still valid. 
However, it will have faster nucleation, in which the subdomain rotation is enhanced because of the dislocation migration in the strained crystal. This will result in the reduced growth rate limited by the dense dislocation structures in the parent grain. 

\begin{figure}
    \centering
    \includegraphics[width=1\linewidth]{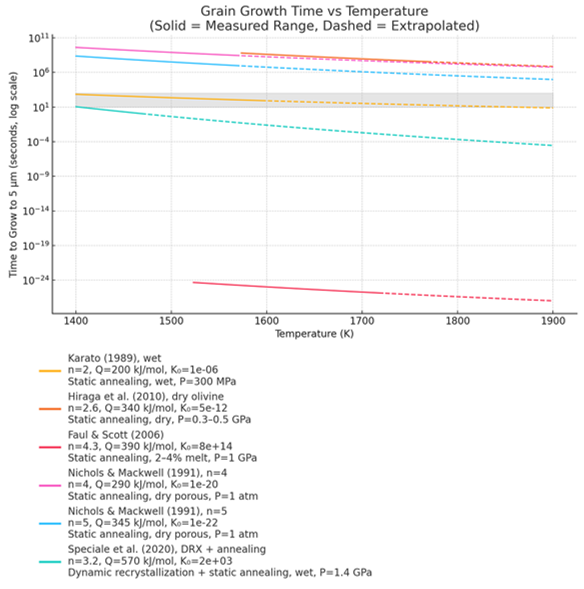}
    \caption{Olivine grain growth time under different experimental conditions}
    \label{figure6}
\end{figure}

Using the available experimental conditions from the literature with changing $n$, $Q$, $K_0$, we extrapolated the time to grow 5 $\si{\mu m}$ with viable temperature from 1200\si{K} up to 1800\si{K} (Fig.~\ref{figure6}), to provide a model of the condition for the Type 1 subgrain formation. As in the figure, the static crystallization in the dry condition, high porosity, and no dynamic driving force will result in the elevated growth components or a small growth factor that greatly slows the growth ($>10^5$/s) even at $1800\si{K}$ \cite{nichols1991grain, hiraga2010grain}. The presence of melt will greatly facilitate subgrain growth \cite{faul2006grain} because of the enhanced subgrain boundary mobility. Similarly, in the dynamic recrystallization, the extra driving force is from the dislocation migration forming subdomains, and the further rotation will form separated grains but with reduced grain size \cite{speciale2020rates}. The model fits our observation for Type 1 subgrains that were formed by a fast nucleation during dynamic recrystallization, resulting in a reduced grain size and very limited grain growth. Shock melt was observed around the parent grain. Therefore, this recrystallization process is possible to be enhanced by presence of the micro-melt interstitially, but it is likely being assisted by the dislocation migration as the additional driving force to facilitate the subdomain rotation. 

The cooling time will affect the size of the crystallite. The shock pulse is ephemeral with the timescale from nanosecond to seconds during impact events, and the duration of the post-shock annealing depending on the rock volume. In the static annealing experiment for olivine shocked to $60\si{GPa}$ \cite{bauer1979experimental}, below $400 ^\circ\mathrm{C}$ or equivalent to $673\si{K}$, there was no change on the texture reported regardless of the annealing time. If we use $673\si{K}$ as the lower limit of the cooling temperature, given the thermal diffusivity of olivine of $10^{-6}\si{m^2/s}$ \cite{fritz_revising_2017}, for a shock-heating annealing starting from $1800\si{K}$, the cooling time is around $2.3\si{s}$ within diameter of 1\si{mm} volume. The short cooling time is plausible for the growth of the Type 1 subdomains. A longer cooling history would facilitate the further growth of Type 1 subdomains beyond what is observed. Therefore, if the shock heating temperature was $1800\si{K}$, then the olivine grain experienced only a very brief heating of several seconds after the shock pulse to retain the observed olivine Type 1 subgrains.

\section{Dark-field X-ray microscopy data}
All dark-field X-ray data were collected at ESRF beamline ID03. The NWA 7721 DFXM datasets were acquired during ESRF sessions by Detlefs et al.~\cite{Detlefs2028_DCT3DXRD_TopoTomo_DFXM_ihma658,Detlefs2027_Shocked_Meteorites_Calcite_ihes145,Detlefs2027_Sample_Tests_Meteorite_CaCO3_ihes143}. The {\AA}heim olivine peridotite dataset was collected during the HERCULES practicals~\cite{Antony2028_HERCULES_Practicals}. We further compiled the KAM and mosaicity data for Type 1, Type 2, and {\AA}heim peridotite into .h5 files and uploaded along with manuscript. These data are available immediately.

\clearpage


\end{document}